\documentclass[referee,a4paper,12pt,traditabstract]{swsc} 
\pdfoutput=1
\usepackage{graphicx}
\usepackage{txfonts}
\usepackage{subfigure}
\usepackage{epstopdf}
\usepackage{lineno}
\usepackage[authoryear,round]{natbib}
\usepackage[hyphens,spaces,obeyspaces]{url}
\usepackage[backref]{hyperref}
\usepackage{xcolor}

\hypersetup{colorlinks=true,citecolor=cyan,urlcolor=cyan,linkcolor=blue}

\begin{document} 

 \title{Solar radio emission as a disturbance of aeronautical radionavigation}
 \author{Christophe Marqu\'e \inst{1}
          \and
	  Karl-Ludwig Klein\inst{2}
	  \and
          Christian Monstein\inst{3}
	  \and \\
          Hermann Opgenoorth\inst{4}
	  \and
          Antti Pulkkinen\inst{5}
	  \and
          Stephan Buchert\inst{4}
	  \and
	  S\"am Krucker\inst{6}
	  \and
	   Rudiger Van Hoof\inst{7}
	   \and
	   Peter Thulesen\inst{8}
	   }

\institute{Solar-Terrestrial Center of Excellence - SIDC, Royal Observatory of Belgium, \\
              Avenue Circulaire 3, B-1180 Brussels, Belgium;
             \email{\href{C. Marqu\'e}{christophe.marque@oma.be}} 
         \and
 	LESIA-UMR 8109  and Station de radioastronomie de Nan\c cay - Observatoire de Paris, CNRS (further affiliations: PSL Res. Univ., Univ. P \& M Curie and Paris-Diderot, University of Orl\'eans, OSUC), 
	F-92190 Meudon, France; 
              \email{\href{K-L. Klein}{ludwig.klein@obspm.fr}} 
         \and
            Institute for Particle Physics and Astrophysics, ETH Z\"urich, Wolfgang-Pauli-Strasse 27, CH-8093 Z\"urich, Switzerland; 
             \email{\href{C. Monstein}{monstein@astro.phys.ethz.ch}}
         \and
            Swedish Institute of Space Physics, Box 537, SE-75121 Uppsala, Sweden 
          \and
             Goddard Space Flight Center Greenbelt, MD, USA 
          \and
             University of Applied Sciences and Arts Northwestern Switzerland, Windisch, Switzerland      
           \and
             Belgocontrol, Tervuursesteenweg 303, B-1820 Steenokkerzeel, Belgium
             \and 
             Air Greenland - Engineering Department, P.O. Box 1012, Nuuk, Greenland
       }
       
       \authorrunning{C. Marqu\'e et al.}
       \titlerunning{Solar radio emission as a disturbance of aeronautical radionavigation}
   \date{}

  \abstract {On November 4th 2015 secondary air traffic control radar was strongly disturbed in Sweden and some other European countries. 
  The disturbances occurred when the radar antennas were pointing at the Sun. In this paper, we show that the disturbances coincided with
   the time of peaks of an exceptionally strong ($\sim 10^5$ Solar Flux Units) solar radio burst in a relatively narrow frequency range around 1~GHz. 
   This indicates that this radio burst is the most probable space weather candidate for explaining the radar disturbances. The dynamic radio spectrum shows that the high flux densities are not due to synchrotron emission of energetic electrons, but to coherent emission processes, which produce a large variety of rapidly varying short bursts (such as pulsations, fiber bursts, and zebra patterns). The radio burst occurs outside the impulsive phase of the associated flare, about 30 minutes after the soft X-ray peak, and it is temporarily associated with fast evolving activity occurring in strong solar magnetic fields. While the relationship with strong magnetic fields and the coherent spectral nature of the radio burst provide hints towards the physical processes which generate such disturbances, we have so far no means to forecast them. Well-calibrated monitoring instruments of whole Sun radio fluxes covering the UHF band could at least provide a real-time identification of the origin of such disturbances, which reports in the literature show to also affect GPS signal reception.}
    

   \maketitle
   
\section{Introduction}
Solar eruptive events such as flares or coronal mass ejections (CMEs) are often accompanied by radio emission produced by supra thermal electrons propagating in open or closed magnetic structures of the solar corona. Solar radio bursts have been observed basically in the whole radio spectrum that current technology makes accessible, from a few hundreds of GHz down to a few kHz. For technical reasons, the VHF-UHF domain ($\sim$ 30--3000~MHz) has been monitored since the early days of solar radio astronomy, and it appeared particularly important for the understanding of fundamental processes occurring during eruptive events \citep{pick2008}. From the meter to decimeter range, characteristic burst signatures correspond to well identified physical processes, such as shock waves (type II bursts), electron beams streaming along open or elongated magnetic field lines (type III bursts) or electron populations trapped in eruptive flux ropes and post-flare loops (type IV bursts). Type IV bursts are oftern accompanied by long duration events observed at EUV or soft X-ray wavelengths, and coronal mass ejections. Discovered in the 1950s in the meter range \citep{boischot1958}, they have been since then observed from the decimeter to the decameter range. They often consist of spectral fine structures (broadband pulsations, zebra bursts, fiber bursts) superimposed on a continuum emission (for recent reviews see \citet{chernov2006} or \citet{nindos2008}). Finally, these bursts often show high brightness temperature and flux density, which represents a challenge in terms of emission mechanisms \citep{cliver2011}.

The potential impact of such intense radio emission on technologies relying on the use of the radio spectrum has been an important topic of research in the field of space weather \citep{bala2002}. Several technologies have been impacted by solar radio bursts occurring in the VHF - UHF bands: Global Navigation Satellite Systems (GNSS), where the solar emission affects the carrier-to-noise ratio of ground based receivers in scientific or industrial applications \citep{klobuchar1999, cerruti2006, demyanov2012}, and radar systems used in military applications for the airspace surveillance \citep{knipp2016}. It is also worth remembering that one of the first published reports of radio emission coming from the Sun concerns the jamming of military radars during World War II \citep{hey1946}.

In this paper, we report on an intense solar radio burst that occurred on November 4th 2015, when air traffic radars were disrupted in Sweden, and milder disturbances were observed elsewhere in Europe. A short account of the radar disturbances is presented in Sect.~\ref{Sec_Rad} to characterise the timing of the incidents. In Sect.~\ref{Sec_ME} possible space weather causes are explored, including activity in the Earth's magnetosphere and at the Sun. The timing of the disturbances is shown to coincide with strong solar radio emission in the vicinity of 1000~MHz. The spectral characteristics of the radio burst are presented, together with context observations of the associated eruptive activity. The occurence of such events and possible physical causes are discussed in Sect.~\ref{Sec_Dis}, where recommendations for future observations are also formulated.

\section{Disturbances of air traffic control radar on November 4th 2015}
\label{Sec_Rad}
\subsection{Air Traffic Control}
In the following, we outline the main aspects of Air Traffic Control (ATC) radar systems relevant to the present study. Most of the information comes from the   \emph{Guidance Material on Comparison of Surveillance Technologies (GMST)} document published by the International Civil Aviation Organization \citep{icao2007}.

Several radar systems are used to locate and gather information about aircraft in a given air space. Primary Surveillance Radars (PSR) provide basic information about azimuth and distance of a plane, by measuring the time difference between the emission of a pulse and the reception of the signal reflected on the frame of the aircraft. PSR are primarily set up at or near airports and generally operated worldwide either in L band (1215--1350~MHz) or S band (2700--3100~MHz).

Secondary Surveillance Radars (SSR) send coded queries to transponders aboard aircraft and get in return ancillary information on the plane: identification, barometric altitude of the aircraft and for some systems, selected technical parameters. They are "blind" to air planes having no functioning transponders. Two technologies are currently in use: mode A/C radars are an older generation of devices that send queries for name identification (mode A) and barometric altitudes (mode C) of the aircraft. These older radar systems present some limitation in terms of information being transmitted back by the aircraft and are essentially querying constantly all airplanes within their field of view. They are progressively superseded by mode-S radars, which can gather more information about the aircraft and interrogate specific planes only. Both generations of SSR operate in two bands: 1030~MHz for the uplink and 1090~MHz for the downlink; they have a range of about 200 to 250~Nautical Miles (370 to 463~km). 

Other aeronautical services make use of the radio spectrum to help pilots operate the aircraft at critical times. For example, when a plane is about to land in difficult weather conditions (poor visibility), pilots may rely on an Instrument Landing System (ILS) to help them align correctly with the landing path and descent at the appropriate angle and speed towards the ground. ILS relies on VHF bands and provides via different subsystems the deviation in azimuth around the end point of the landing path (the localizer), and the position in elevation above the reference point (the gliding slope) in addition to other ancillary information \citep{faa2012}.

\subsection{Air traffic disturbances}
On November 4th 2015, several European air traffic authorities reported issues with operations of secondary air traffic radar systems at or near local sunset. In the following, we present problems affecting radars in Belgium and Sweden that were made available to us via a post event technical report analysis (Belgium), or that we compiled from several publicly available sources (Sweden). We also report on a possible ILS problem at the Thule Airbase in Greenland.  

\begin{figure*}
\centering
\includegraphics[width=\columnwidth]{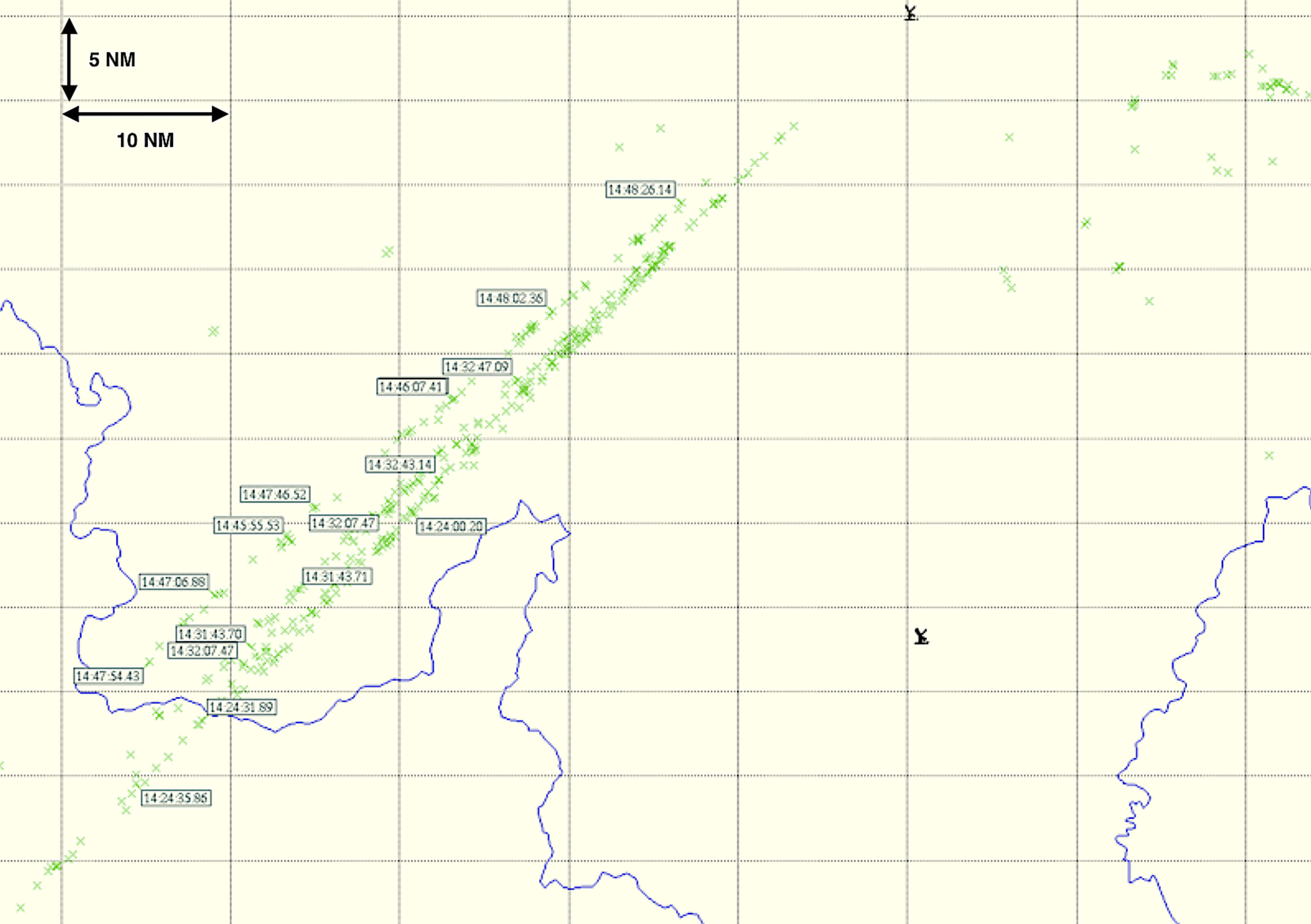}
\caption{\small False echoes (in green) observed in the direction of the Sun at a Belgian A/C radar station, located to the top right. The border with France and Luxembourg appears in blue. The scale of the map, in Nautical Miles is indicated at the top left.}  
\label{fig_radar}
\end{figure*}

\subsubsection{Disturbances in Belgium}
\label{belgium}
Belgocontrol, the air traffic authority in Belgium, reported issues with a secondary A/C radar. False echoes, representing non-existing planes,  were observed only in the direction of the Sun during two periods of time : 14:19:23--14:33:35~UT and 14:45:52--14:49:18~UT. Figure \ref{fig_radar} shows the false plane positions aligned in the direction of the Sun observed at that time. No traffic interruption occurred due to this incident as the control software filtered out correctly the false information.

\subsubsection{Disturbances in Sweden}
\label{sweden}
Severe disruptions of the air traffic over the southern part of Sweden occurred on that day resulting in a {\it de facto} partial closure of the airspace and delayed arrivals and departures according to reports in the media \citep{tloc}. Based on publicly available information from the Swedish air traffic authority (Luftfartsverket; hereafter LFV), ATC secondary surveillance radar systems could not display proper information to the air traffic controllers, which lead the authorities to reduce aircraft movements for safety reasons \citep{lfv2015}. More details were later made public by LFV during  a workshop in November 2017, in Uppsala (A. J. Andersson, 2017; personal communication) indicating for this event that a simultaneous series of ATC radar disturbances occurred at several sites in Sweden starting around 14:19~UT. These disturbances were observed as ghost echoes in straight lines in the direction of the Sun and took place first between 14:19 and 14:34~UT, and then later on between 14:47~UT and 14:50~UT. The disturbances were more pronounced during the first time period. The magnitude of disturbances resulted in some stations being overloaded and triggering loss or delays of tracks. 

 Media reports and an early on press release pointed at a space weather event as the cause of the disruptions, evoking, confusingly a solar flare and a geomagnetic storm \citep{tloc}. 

\subsubsection{Other mentions of air traffic disturbances}
Disturbances affecting SSR systems have been recently made public by air traffic authorities from Norway (Avinor Flysikring) during the Birkeland Anniversary conference 2017 conference, in Norway (A. D. Skjervold, 2017; personal communication\footnote{\url{http://www.mn.uio.no/english/about/news-and-events/events/The Birkeland Anniversary/andreas-d-skjervold_150617.pdf}}). Like in the Belgian and Swedish cases, ghost or false signals lining up in the direction of the Sun around 14:30~UT, near the peak of the event, were observed at several ATC radar stations in Norway without resulting in traffic perturbations. The same reference indicates that older generation radars (A/C types) were more "impacted".

\subsection{Disturbances in Greenland}
\label{greenland}
On November 4th 2015, an Air Greenland  plane landing at Thule Airbase in Greenland at 14:49~UT experienced technical issues above 4000~feet altitude with a conflicting report between an ILS localizer (at 109.5~MHz) indicating a correct alignment with the runway and the autopilot being unable to hold on that same position information. The landing went without further complications. After the flight the ILS equipments were checked for malfunction but were cleared of any defect.

\section{Solar and magnetospheric activity on November 4th 2015}
\label{Sec_ME}
In the search of a candidate space weather event to explain the radar disturbances, we consider the timing of activity in the Earth magnetosphere and in the solar corona.

Sustained geomagnetic activity is revealed by the Kp index on and around November 4th. Kp exceeds values of 5, the lower limit for the NOAA qualification\footnote{\url{http://www.swpc.noaa.gov/noaa-scales-explanation}}  as a ``minor" geomagnetic storm, on 3rd, 4th, and 7th of November. But Kp stays below 7, the threshold of the ``strong" category. Following the NOAA classification 1700 minor geomagnetic storms occur during a solar activity cycle, and 200 strong ones. There is hence no exceptional geomagnetic disturbance that could be a candidate space weather event explaining the air traffic radar problems on 4th November.

\subsection{Solar eruptive activity}
A solar flare (SOL2015-11-04T13:52) took place in NOAA AR 12443 while close to the central meridian. This region began its transit on the solar disk on October 29th as a complex sunspot group, with a McIntosh classification evolving from Ehc on October 30th to Fkc on November 4th. On the day of the flare, its magnetic type, according to the Mount-Wilson classification was $\beta \delta$. From October 29th till November 4th noon, an intense flaring activity took place with two strong B class flares, 55 C class flares, and one M class flare. None of these flares was accompanied by intense broadband radio emissions. Similarly, no significant noise storm activity was recorded at metric wavelength during that time frame. 

SOL2015-11-04T13:52 was a long duration soft X-ray M class flare, also observed in H$\alpha$ by the Kanzelh\"ohe Observatory as an 1B optical flare. It was accompanied by a halo CME first observed at 15:12 UT in the field of view of the SoHO/LASCO-C2 coronagraph \citep{brueckner1995}. A streamer deflection suggests a CME-driven shock wave, which is indeed confirmed by metric and decametric type II burst emissions.

In the following we examine in more detail the soft X-ray and radio observations of the event.

\subsubsection{Instrumentation}
We make use of soft X-ray observations provided by the XRS instrument \citep{garcia1994} aboard the \emph{Geostationary Operational Environmental Satellites} (GOES-13 and GOES-15), and EUV contextual images from the AIA instrument aboard the Solar Dynamics Observatory spacecraft \citep{lemen2012}. 

The radio spectra discussed in this paper were collected at three facilities in Belgium (Humain radio astronomy station), France (Nan\c cay radio astronomy station), and Switzerland (Bleien and Z\"urich Observatories). 

In Humain (Belgium), a 6-m dish radio telescope is used to monitor the solar activity on a daily basis. A dual-polarised log-periodic feed antenna covers the band 300--3000~MHz, but currently the setup is such that flux density measurements are performed only between 275 and 1495~MHz (in particular, no polarisation information is available). The Humain Solar Radio Spectrometer (HSRS) attached to the antenna is a Software Defined Radio receiver which is scanning the whole band in steps of 20~MHz. For each step, a Fast Fourier Transform spectrum is calculated and stacked with the others to produce a wide band spectrum with a constant frequency resolution of $\sim$ 98~kHz and a time resolution of $\sim$ 0.250~s.

The ORFEES (\textit{Observations Radio pour FEDOME et lÕ'\'Etude des \'Eruptions Solaires}) spectrograph setup in Nan\c cay observes the flux density of the entire Sun between 100 and 1000~MHz, using a parabolic dish of diameter 5~m and a log-periodic dipole array. The spectral range is covered by five bands of about 200~MHz bandwidth. Each band is scanned within 0.1~s. The observations were complemented towards lower frequencies (20--80~MHz) by the Nan\c cay Decameter Array \citep[NDA;][]{lecacheux2000}, a sensitive spectrograph connected to a field of 144 helicoidal antennas.

 While routine solar radio observations were interrupted recently in Switzerland, the event under study has fortunately been observed by two instruments: one in Bleien, a Callisto spectrometer \citep{benz2005b} connected to a cylindrical horn-antenna covering the band 1000--1300~MHz, and mounted on a 5-m dish telescope, and the other one in Z\"urich: a Callisto receiver connected to a cylindrical horn antenna at 1427~MHz, mounted on a 5-m dish telescope. In both cases, heterodyne down-converters were used to accommodate for the original input frequency range of the Callisto receivers (45--870~MHz).

In addition to dynamic spectra, we also used fixed-frequency time histories observed by the Sagamore Hill station of the {\it Radio Solar Telescope Network} (RSTN; \citealp{kennewell2008}) operated by the US Air Force.  

\begin{figure}[htbp]
\centering
\includegraphics[width=0.7\columnwidth]{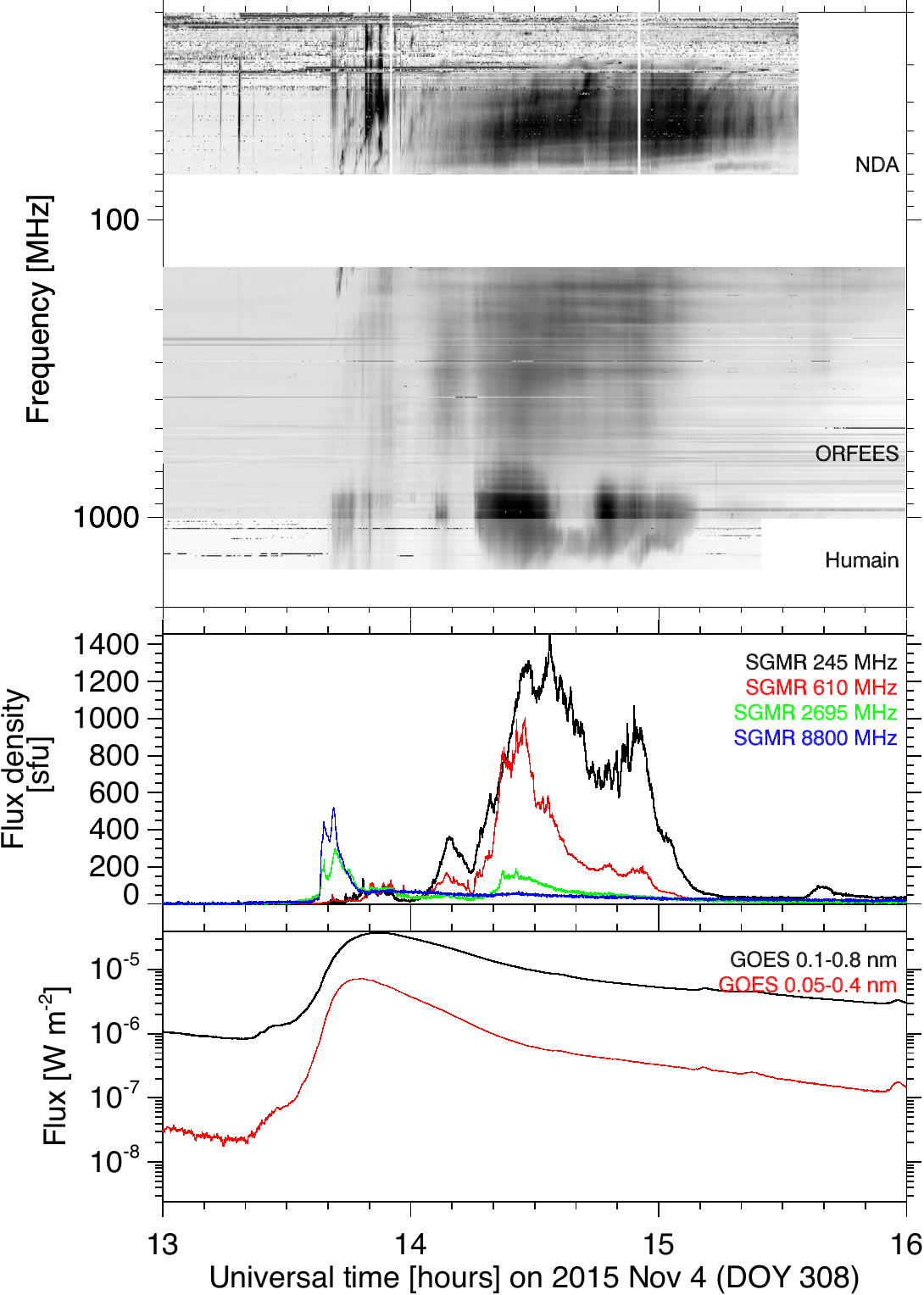}
   \caption{Overview of the November 4th 2015 event as observed in soft X-rays (GOES, bottom panel), at selected radio frequencies (Sagamore Hill station of RSTN, 2nd from bottom) and as dynamic spectrograms (HSRS spectrograph in Humain, Belgium; ORFEES spectrograph and Decameter Array, Nan\c{c}ay, France). Dark shading shows strong emission. Horizontal lines are from terrestrial emitters.}
   \label{fig_sum}
\end{figure}

\subsubsection{Soft X-ray and radio emission}
Figure~\ref{fig_sum} shows the time histories of the full Sun soft X-ray (bottom panel) and radio emission. The radio spectra cover a large band including the radar frequencies. The event looks unremarkable in GOES and in radio bands monitored by the RSTN Sagamore Hill station. The soft X-ray burst started at 13:35~UT and peaked at the M3.7 level at 13:52~UT. This burst is significant, but moderate: judging from the time since the beginning of the GOES soft X-ray observations, 31 such bursts or stronger are expected to occur per year. Moreover the burst had its most intense phase well before the radar incidents. An impulsive microwave burst was observed during the soft X-ray rise (second panel from bottom, 2695 and 8800~MHz), with faint counterparts at frequencies below 1000~MHz. This is a typical impulsive burst, due to gyrosynchrotron emission of mildly relativistic electrons. Hard X-ray emission was observed up to 100~keV by the {\it Fermi} mission\footnote{\url{http://sprg.ssl.berkeley.edu/~tohban/browser/}}.

While the soft X-rays and the microwaves decayed, a new rise to much higher flux density than during the impulsive phase occurred between about 14:05 and 15:10~UT at frequencies below 1500~MHz.  The emission was broadband, with a low-frequency limit at or below 30~MHz. This limit is affected by the decrease of sensitivity of the Nan\c cay Decameter Array antennas towards low frequencies, the emission probably extends to still lower frequencies. The large bandwidth and long duration show that this is a type~IV radio burst.

The flux densities measured by RSTN are moderate, with peak values not exceeding 1400 Solar Flux Units (SFU\footnote{$1~\mathrm{SFU}=10^{-22}\mathrm{W}\cdot\mathrm{m}^{-2}\cdot\mathrm{Hz}^{-1}$}) in any band monitored by this network but at 1415~MHz. The type IV burst is by no means a strong radio event, with the marked exception of the frequency range 800--1500~MHz seen in the HSRS, ORFEES and Callisto dynamic spectra. The peak flux density at 1415~MHz measured by RSTN is about 5800~SFU. Between 1000 and $\sim 1200$~MHz, two periods of particularly strong emission occur at 14:15--14:30 and 14:45--15:00~UT. This timing is in remarkable agreement with the strongest radar disturbances ($\sim$14:19--14:34 \& 14:46--14:49~UT in Belgium), and points to the solar radio burst as the cause of these disturbances.

\subsection{Calibration of the radio flux densities}
As shown in Fig. \ref{fig_sum}, the peak time of the light curves corresponds to the development of the type IV burst emission as observed by ORFEES and HSRS, and in particular to the strengthening of the high frequency part of the Type IV burst (above $\sim$900~MHz). Even uncalibrated measurements provided by Nan\c cay, Humain and Bleien observations hinted at a much stronger emission peak in that high frequency range. We performed a cross-calibration procedure based on relative measurements with respect to the quiet Sun level observed prior to that event. A spectrum of the quiet Sun level has first been calculated by fitting the Quiet Sun level observations provided by the Nobeyama Radio Polarimeters \citep{torri1979}, the US-Air Force RSTN network and the F10.7 measurements at Penticton \citep{tapping2013}. 

The fitting was performed using a Gaussian Processes regression \citep{pedregosa2011} to cope with the large spread of RSTN values.  For each frequency, we calculated a daily average and its associated standard deviation. For Nobeyama, we assumed a 1\% standard deviation. The Quiet Sun flux density in the mid of the radar band (1060~MHz) is found to be 75~SFU. This is the value used in our cross-calibration. As no polarisation measurements are available in the radar band, we assumed an unpolarised emission.

\subsubsection{ORFEES calibration}
The ORFEES spectrograph upper limit (1000~MHz) falls short of the radar band but is close enough to provide pertinent information on the flux density. The antenna was off-pointed by about half a beam at 1000~MHz since about 10:20~UT. The flux density was recalibrated at 1000~MHz using the following procedure: the flux densities measured in the morning, before the antenna was pointed at the Sun, are considered to represent the sum of the cold sky and contributions from the hardware (antenna, receiver, cables). Their average value was subtracted from the flux density time profile. It was then assumed that, after this subtraction, the flux density between 12:10 and 12:20~UT represented the quiet-Sun level. The Quiet Sun flux density inferred from the fit of the quiet Sun spectrum just mentioned was used to evaluate the calibrated flux density at 1000.11~MHz. As a result, we find that the 1000~MHz emission during the burst on November 4th 2015, between 13:30 and 15:30~UT, reached a peak value of $10^5$~SFU ($0.98 \cdot 10^5$~SFU at 14:28:12~UT).

\subsubsection{HSRS Calibration}
In Humain, the two radar bands 1030 and 1090~MHz are effectively observed by the instrument, but due to the proximity of an ATC radar (Saint Hubert Airport) about 21~kilometres to the South East of the Humain station, the light curves in these two bands are heavily disturbed by the radar emission. For this event a light curve at 1060~MHz (central frequency) was calculated as the median, in spectral dimension, of observations between 1030 and 1090~MHz. 

\begin{figure}
\centering
\includegraphics[width=0.7\columnwidth]{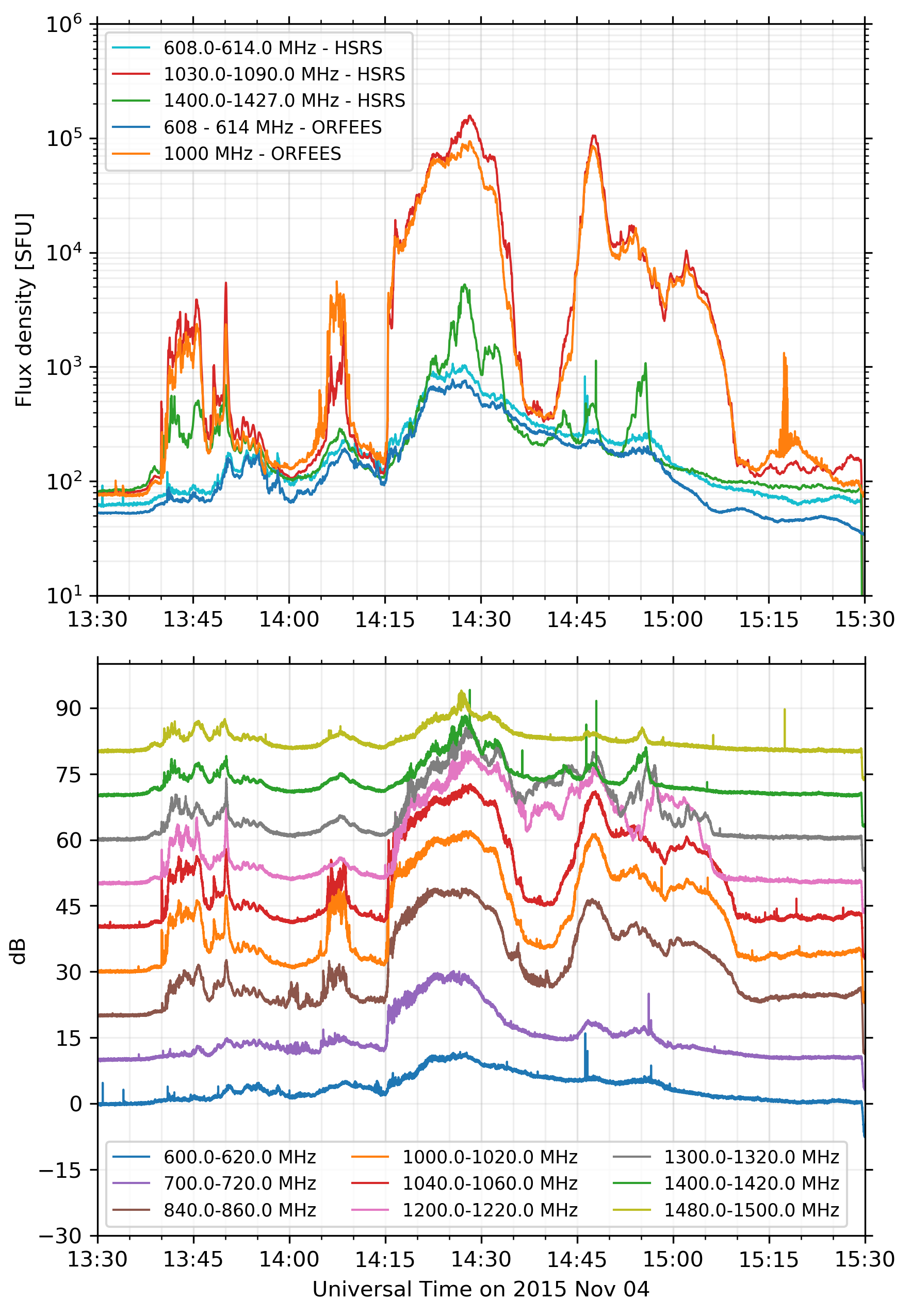}
\caption{Top: light curves in SFU of the November 4th 2015 event from the HSRS and ORFEES spectrographs. Bottom: light curves in relative dB from the HSRS spectrograph at several frequencies between 600 and 1500~MHz.}
\label{fig_lc}
\end{figure}

Before the raw data were scaled to SFU, estimates of the sky and ground contributions were made. Figure~\ref{fig_lc}, top panel, shows the light curves recorded at Humain between 13:30 and 15:30~UT at 1060~MHz and two other bands which are registered in Belgium for Radioastronomy, 608--614~MHz and 1400--1427~MHz. Observations stopped shortly before 15:30 UT, about 45~min before local sunset, when the telescope went back automatically to its night time "parking position". For all these bands, quiet sun flux densities were derived from the modelled quiet sun spectrum mentioned earlier. For comparison, light curves observed by ORFEES in the band 608--614~MHz and at 1000~MHz are also shown. For HSRS, the peak flux density at 1060~MHz (band 1030 - 1090~MHz) is approximately 157~kSFU at 14:28:18~UT, with two time intervals where the flux is above 100~kSFU: 14:26:12--14:29:33 and  14:47:26--14:47:51~UT.  The peak flux values in the two other bands are 1034 and 5245~SFU at 608--614 and 1400--1427~MHz, which can be compared to the peak fluxes reported by NOAA in the same bands (1000 and 5800~SFU at 610 and 1415~MHz respectively).

\subsubsection{Swiss observations calibrations}
The calibration of the Callisto observations in Bleien and Z\"urich followed the same procedure as for the two other radio data sets, and made use of the quiet sun spectrum aforementioned to derive a calibrated event spectrum for each instrument. 
The Bleien spectrum shows some signs of saturation as the upper limit of the receiver dynamic range is reached. Taking this into account, plus the horn frequency characteristics, the best estimate for the peak flux density value is $1.23\cdot10^5$~SFU over the frequency range 1000--1250~MHz. Peak values in the 1427~MHz band are rather similar in Humain, Z\"urich and RSTN observations (5245, 6300, 5800~SFU respectively).

\subsubsection{Summary of radio calibrations}
Table \ref{table_flux_event} gives an overview of the peak flux values estimated from the three radio observations used in the present study. Within the approximations and known sources of uncertainty, the peak of emission in the radar bands is higher than $10^5$~SFU and possibly as high as $\sim1.6\cdot10^5$~SFU.

\begin{table*}[t]
\caption{Summary of the magnitude of the radio burst (peak flux) at several frequencies}
\label{table_flux_event}
\centering
\begin{tabular}{cccccc}
\hline
\hline
Frequency & ORFEES & HSRS & Callisto & Callisto &NOAA report\\
(MHz)& (Nan\c cay) & (Humain) & (Bleien) & Z\"urich & \\
\hline
610 & 820 & 1034 & & & 1000\tablefootmark{a}\\
1000 & $10^5$\tablefootmark{b} & & & & \\
1060 & & $1.57\cdot10^5$\tablefootmark{c} & & & \\
1000--1250 & & & $1.23\cdot10^5$ & & \\
1415, 1427 & & 5245\tablefootmark{d} & & 6310 & 5800\tablefootmark{e}\\
\hline
\end{tabular}
\\
\tablefoottext{a}{at 14:27~UT}
\tablefoottext{b}{at 14:28:12~UT}
\tablefoottext{c}{at 14:28:18~UT}
\tablefoottext{d}{at 14:27:25-30~UT}
\tablefoottext{e}{at 14:27~UT}
\end{table*}%

\subsection{Spectrum and fine structure of the radio emission in the 600--1500 MHz range}
The time history of the strong radio emission around 1000~MHz is shown in more detail in Figure~\ref{fig_lc}, bottom panel, at selected frequencies across the spectrum. The time histories at the lower (600--620~MHz) and upper (1400--1420 and 1480--1500~MHz) limits of the plotted range are similar, with a rather smooth rise starting 14:15~UT and a maximum near 14:27~UT, which is followed by a decay with superposed similar fluctuations. The emission is clearly broadband. In the frequency range between these limits, in the range 700--1220~MHz, the emission is stronger. In addition, the initial rise is much steeper than at the lower and upper limit of the frequency range. After the steep initial rise, however, the time profiles of the strongest emission again resemble those at the lower and higher frequencies. They have a similar slow further rise, and peak at similar times. So the intense emission in the range 700-1220~MHz is superposed on a broadband continuum.

The steep rise is not an instrumental artefact, but a solar phenomenon, as seen by a comparison between the ORFEES and HSRS time profiles at respectively 1000 and 1060~MHz shown in the top panel of Figure~\ref{fig_lc}. It is a second characteristic property of the radio burst spectrum around 1000~MHz, besides its high flux density. 

\begin{figure*}[htbp]
\centering
   \includegraphics[width=\columnwidth]{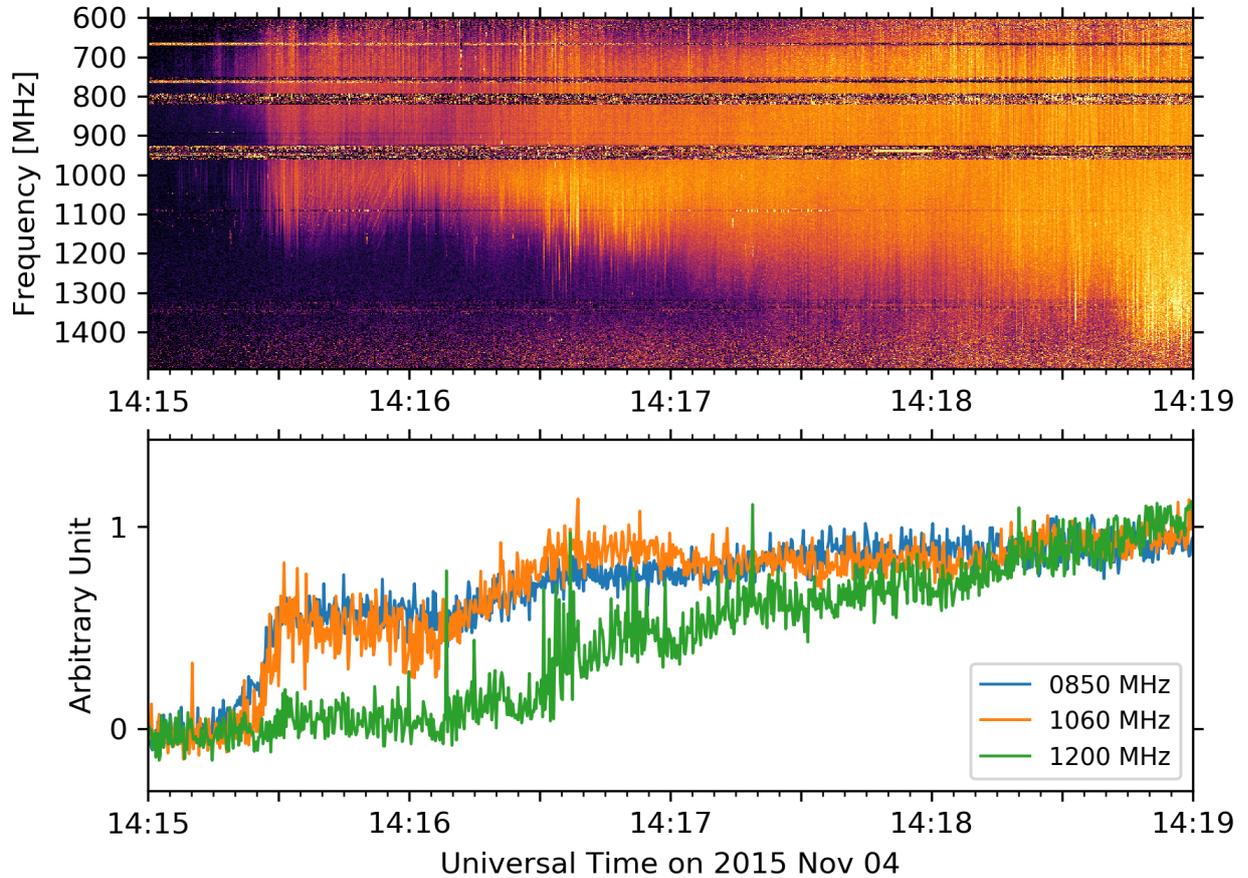}
   \caption{Fine structure of the radio spectrum in the first minutes of the strong emission between 14:15 and 15:00 UT.}
   \label{fig_fsr1}
   \end{figure*}

\begin{figure*}[htbp]
\centering
   \includegraphics[width=\columnwidth]{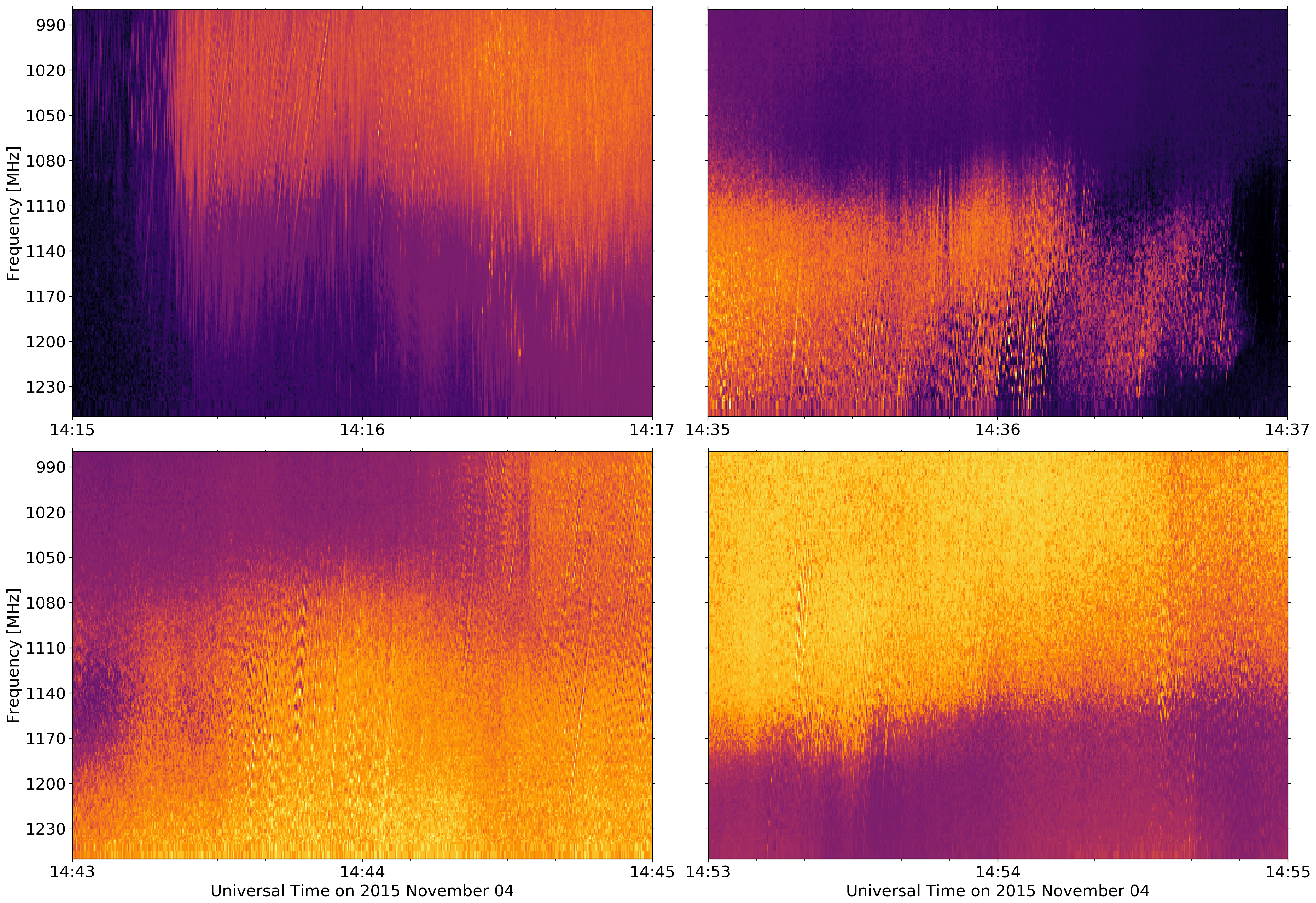}
   \caption{Fine structures consisting of fiber bursts and zebra pattern at different stages of the event as observed by the Callisto spectrometer in Bleien.}
   \label{fig_bleien}
\end{figure*}

\begin{figure}[htbp]
\centering
    \includegraphics[width=\columnwidth]{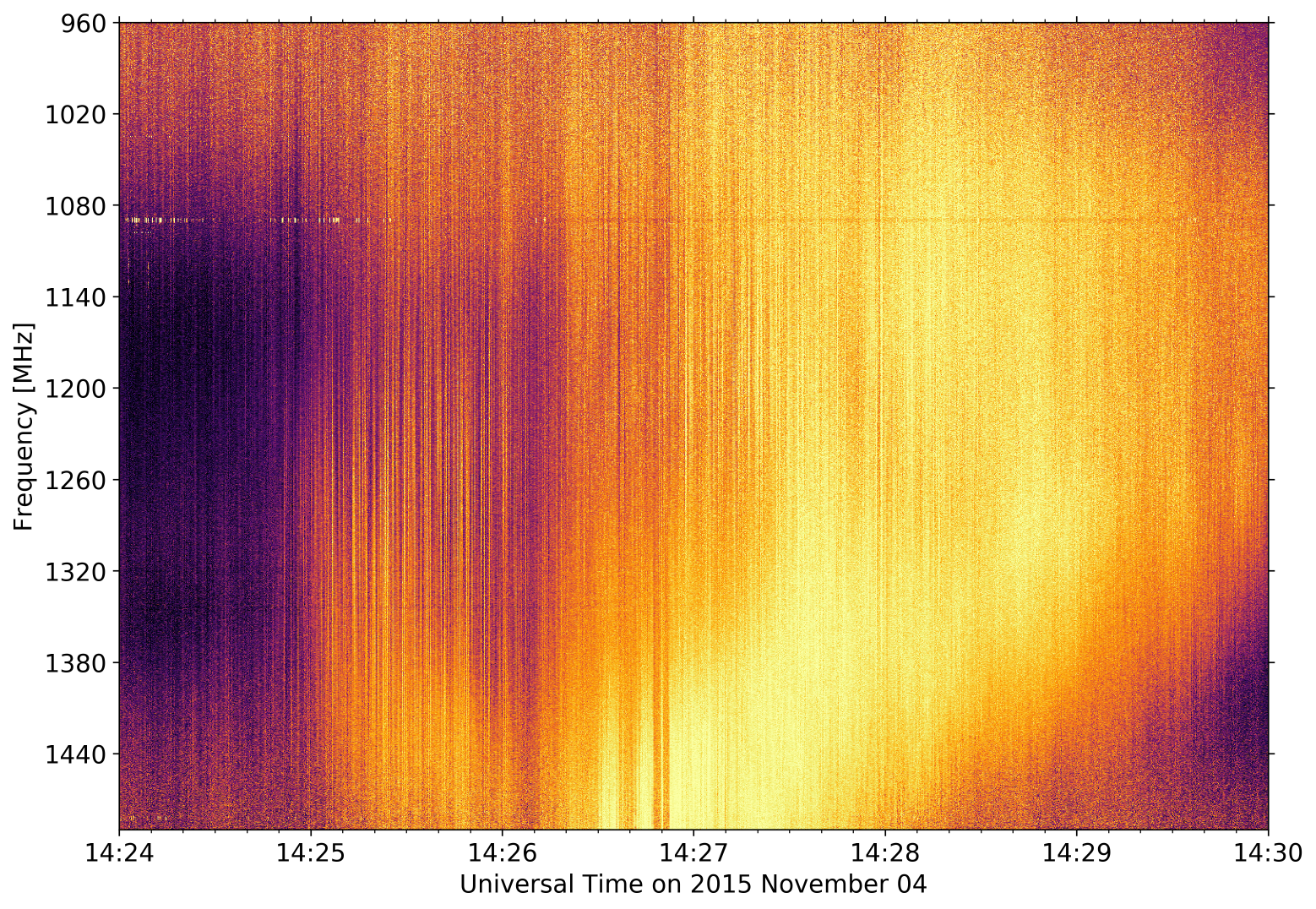}
   \caption{Dynamic radio spectrum observed with HSRS around the peak of the strong emission. Notes that the flux density scale was set independently for  the individual frequencies. The apparent maximum near 1400~MHz is an artifact of this procedure. The horizontal line at 1090~MHz, is actually the downlink part of a secondary ATC radar transmission close the station of Humain.}
   \label{fig_fsr2}
\end{figure}

More detailed spectra of the strong burst in the range 600--1500~MHz are shown in Figures~\ref{fig_fsr1} to \ref{fig_fsr2}. The spectra have been processed to increase the visibility of the fine structures. During this process the relative intensity from one frequency band to the next is lost. The spectrum in Figure~\ref{fig_fsr1} comprises broadband pulsations at the beginning of the first intense event, 14:15:30~UT, in the range 600--1150~MHz. Comparison of time profiles in fixed-frequency ranges plotted below the dynamic spectrum shows that it is indeed the pulsations which create the steep rise of the flux densities noted above. The 250~ms time evolution of HSRS probably does not resolve individual pulsations when they are too densely packed. We cannot rule out that the intensity observed during that flare arises from numerous spike bursts as observed during the 2006 December 06 event \citep{cliver2011}.

The visual impression from the spectrum in Figure~\ref{fig_fsr1} is that the bright emission starts with individual pulses near 14:15:15, which thereafter become brighter and more densely packed. In the centre of the band the spectrum shows a more or less continuous emission after some time, but the succession of pulsations is clearly visible at the high-frequency limit throughout the spectral plot. So one may suspect that the intense emission is a supposition of many broadband pulsations. Other types of spectral fine structure become visible: fiber bursts (for instance, 14:15:40--14:16:00 between 960 and 1200~MHz), zebra pattern  ({\it e.g.}, 14:15:30--14:15:40, same frequency range). These types of structures are observed over the whole event as shown in Figure~\ref{fig_bleien}. These fine structures are discussed in \citet{isliker1994, jiricka2001, aurass2003, chernov2006}. 

The dynamic spectrum during the maximum of the radio burst near 1000~MHz is shown in Figure~\ref{fig_fsr2}. Again broadband pulsations are clearly discernible.
The spectral fine structures identified above are all well known from the observations of type~IV bursts at frequencies of several hundreds of MHz. They are usually ascribed to micro instabilities of non thermal electron populations trapped in coronal magnetic fields. A sustained feature in the range of the radar frequencies are the broadband pulsations.

\subsection{The eruptive event: EUV and coronographic observations}
The impulsive phase of the flare between $\sim$ 13:25 and 13:50 UT is associated with several signatures in the EUV range: expanding loops, a coronal wave and the appearance of a dimming region to the North of the active region. A partial halo CME going essentially southward appears in the LASCO C2 field of view at 14:24 UT. The gradual phase of the flare, which overlaps with the radio event, shows classical post flare loops in the core of the active region. The time of the radio event corresponds to a late phase in this post-flare loop period, with new loops forming to the North-West of the active region. 

\begin{figure}[htbp]
 \centering
   \includegraphics[width=\columnwidth]{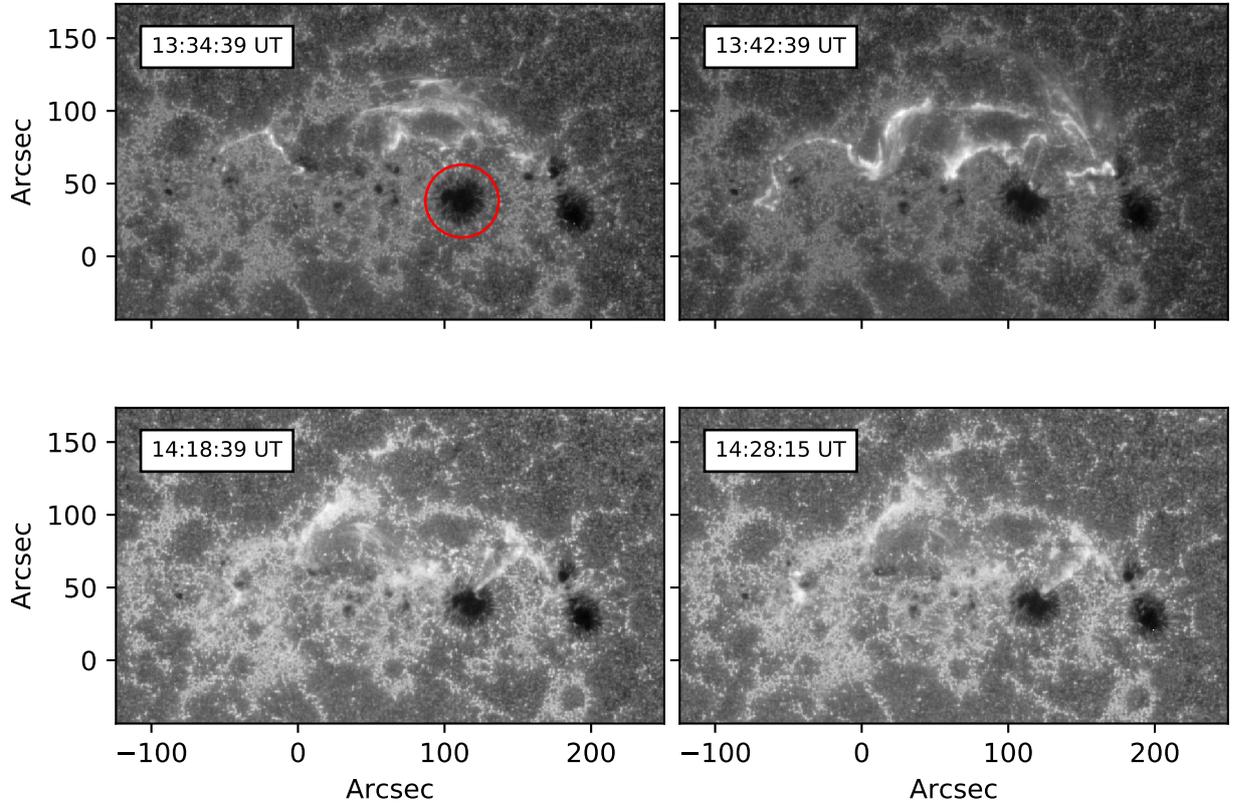}
   \caption{AIA observations at 1600 Angstr\"om at different times during the flaring event. The two panels to the bottom show a bright ribbon sweeping through the umbra of the sunspot circled in red, close to the time of the sharp increase of the radio flux density.}
   \label{fig_aia}
\end{figure}

The penumbra and umbra of the trailing sunspot (see Figure \ref{fig_aia}) are swept over by a late expanding flare ribbon during at least the duration of the strongest and sudden radio flux increase near 1000~MHz. The radio sudden brightening starts slightly earlier (14:15:30~UT) that the beginning of this flare ribbon expansion (14:17~UT). The penumbra appears profoundly modified by this intrusion of the flare ribbon, at least up to the end of the radio event. Earlier during the event, the first and weaker radio flux increase (from 13:40 till $\sim$13:49 UT) corresponds also to a flare ribbon expanding into the same sunspot umbra.

\section{Discussion and conclusion}
\label{Sec_Dis}

The radar disturbances, which in Belgium, Sweden and Norway consisted of false echoes received when the antennas looked sunwards, had no coincident geomagnetic activity enhancement. But the striking temporal association between the two periods of strongest disturbances and the two peaks of extreme solar radio emission in a band covering the radar frequencies points at the Sun as the origin of the perturbations.

\subsection{Impact of the solar burst on radar systems and radionavigation}

Detailed technical information concerning secondary radars is not publicly available. However, compatibility studies for the coexistence of different services in L band mention generic information about radar susceptibilities to interferences. Table \ref{table_radar_interference} summarises basic facts extracted from  Eurocontrol compatibility studies  \citep{frequentis2007,micallef2009} and private communications (B. Collard, DSNA \& L. Lap\`ene, DGAC).

\begin{table*}[!h]
\caption{Parameters for the calculation of radar sensitivity to interferences\tablefootmark{a}}
\label{table_radar_interference}
\centering
\begin{tabular}{ccc}
\hline
\hline
Parameter & Description & Value\\
\hline
$G$ & Maximum antenna gain & 27~dBi \\
PT & Pointing in elevation & $\sim$ 6--8$\degr$\\
BW & Beam width & $10\degr$ at 3~dB\\
S$_{min}$ & Minimum detectable signal at antenna output & -103~dBm\\
$CL$ & Cable loss & 2~dB \\ 
$RBW$ & Typical receiver bandwidth & 5.5~MHz\\
S/I & Signal to interference ratio & 12~dB \\
MG & Aeronautical safety margin and multiple technology allowance & 12~dB \\
\hline
\end{tabular}
\\
\tablefoottext{a}{We follow here the methodology by \citet{frequentis2007}, where a conservative interference threshold at receiver input is calculated as: S$_{min}$+G-CL-(S/I)-MG, which amounts to -102~dBm}
\end{table*}%

From these elements it appears that a conservative interference level threshold at a radar receiver input is about $-102~\mathrm{dBm}$ (see Table \ref{table_radar_interference} for details). This can be compared to the quiet sun flux density and the peak flux density during the event. At 1090~MHz, we considered a quiet sun flux density of 75~SFU. The power at receiver level is therefore:

\[
P_{rec}= 10 \log(75 \times 10^{-22} \mathrm{W.m}^{-2}.\mathrm{Hz}^{-1} \times A_{eff} \times RBW ) - CL \approx -101~\mathrm{dBm}
\]
with the effective area:
\[
A_{eff}=\frac{G\lambda^2}{4\pi}\approx 3~\mathrm{m}^2 \; .
\]
\noindent
This power is close to the threshold of interference, but is not detrimental in normal operation. During the peak of the radio event, for which flux densities above 100 kSFU were observed, the power at receiver level amounted to at least $P_{rec}\approx -68~\mathrm{dBm}$, when radars pointed at the Sun. This is about 34~dB above the interference threshold. We can therefore conclude that radars, for which the Sun was between 2--$12\degr$ elevation, could easily pick up the radio burst, at power levels several orders of magnitude above the interference threshold.

We also reported an issue (see section \ref{greenland}) with ILS equipments operating at 109~MHz for a plane landing at Thule Airbase at the end of the radio event (14:49~UT). Post-event discussions with technicians from the company at Thule Air Base indicate that this was unusual and considered as "first time observations". The aircraft suffering the incident had its ILS equipment checked afterwards for any defect. At the time of the event, and at the altitude of the plane when the problem was noticed (above 4000~feet), the Sun was ~2.9 degrees below the horizon. In the VHF band, dynamic spectra of the radio emission (Fig.~\ref{fig_sum}, top panel) show a low frequency extension of the type IV burst, with occasional type III bursts. One was in progress at 14:49~UT and was approximately ten to thirty times stronger than the background as measured by the spectra from the Sagamore Hill and San Vito RSTN stations. 

Refraction is needed in order to make the burst visible to the airplane. At these low frequencies, strong refraction in the ionosphere due  to atmospheric gravity waves is frequently observed, and it goes along with focussing effects that are readily visible in solar radio spectra \citep{mercier1989}. Raytracing calculations performed for Air Greenland for that event seem to corroborate this conclusion (Per H\o eg, 2016, DTU Space, Private Communication). We conclude that the disturbance at 109~MHz can in principle be related to solar radio emission. However,  Fig.~\ref{fig_lc} shows that the burst around 1~GHz, which we identified as the source of the radar disturbances, clearly stands out above the broad band radio event at frequencies above 600~MHz.

\subsection{Mechanism of the radio emission}

Unlike the VHF burst, the radio emission around 1000~MHz is about 2000 times stronger than the Quiet Sun. Explaining such a magnitude remains challenging. The brightest radio emission in the vicinity of the radar frequencies on November 4th 2015 is a collection of spectral fine structures, especially broadband pulsations. The type~IV burst has an overall evolution that is similar over a much more extended spectral range than the bright emission. The fine structure with its sudden onset is a distinctive feature at those frequencies where the burst is bright. In addition, fiber bursts and zebra patterns are observed during some time intervals. They are ascribed to the interaction between upper hybrid waves and whistlers, which are generated in microinstabilities of trapped non thermal electron populations \citep{zlotnik2013}. In the corona, where the plasma frequency is usually believed to exceed the cyclotron frequency, such instabilities create plasma waves, which then couple to radio waves that can escape.

An alternative process that creates directly escaping radio waves, and is known as the mechanism of very intense radio bursts in planetary magnetospheres, is the electron cyclotron maser. This mechanism is usually expected to not operate in the solar corona, because of the high electron plasma frequency. The exceptional occurrence of the strong radio burst near 1000~MHz could of course be explained by an exceptional situation of plasma parameters in the corona, so that electron cyclotron maser emission might arise. The argument was put forward by \citet{regnier2015} and \citet{cliver2011}. However, the identification of well-known fine structures, like fiber bursts and zebra patterns, that are observed in type IV bursts, but usually at lower frequencies where it is still more unlikely that the cyclotron frequency exceed the plasma frequency, makes such an exceptional situation improbable. Instead, it rather argues for coherent plasma emission.

The sudden rise in flux density observed first around 13:40~UT and especially later around 14:15~UT corresponds approximately in time to the expansion of one of the flare ribbons through the umbra of one of the sunspots of the active region (13:40 and 14:17~UT). This behaviour is observed in a surprisingly small number of M and X class  flares. In a survey of 588 M and X class flares observed by the TRACE and Hinode/SOT telescopes, \citet{leping2009} have noticed only 20 events displaying a flare ribbon sweeping across the umbra of one of the two main sunspots of the active region. However, only two of their 20 events are associated with intense solar radio bursts (flux density above 100\,000~SFU) and both occurred in the same region (NOAA AR 10930, on December 6th and 13th 2006). We checked a more recent event, which took place in September 24th 2011, with a peak flux of 110\,000 SFU at 1415~MHz. It also shows an abrupt intensification of the flux density approximately as a flare ribbon sweeps across the umbra of one of the sunspots in the flaring active region. One physical implication could be that particularly strong magnetic fields are involved during the bursts of interest, which in turn could be a new support for the interpretation by a cyclotron maser mechanism. However, at this stage, one cannot draw any firm conclusion linking these two phenomena. 

\subsection{The occurrence frequency of strong solar radio bursts near 1000 MHz}

A question that arises from this analysis is how frequently such intense radio emission can trigger noticeable or damaging consequences on human technologies. \citet{nita2002}  analysed the occurrence of strong radio bursts over nearly 40 years, deriving  the distribution function of burst occurrence depending on frequency, flux and phase of the solar cycle. Following their analysis, events of magnitude greater than $10^5$~SFU in the band 1 to 1.7~GHz occur every $\sim8$~months during solar maximum, every $\sim38$~months during solar minimum and every $\sim$~23 months over the 40 years of the survey. That does not necessarily mean that a noticeable effect will be observed at these frequencies. At a given location on Earth, air traffic radar systems will be most affected at or near Sun rise or Sun set. If we suppose that radar systems are vulnerable within a 2~hours window at dawn or dusk, the occurrence rate will be multiplied by $\sim$6: every $\sim48$~months, every $\sim228$~months (19 years) and every $\sim$~138 months (11.5 years) for solar maximum, solar minimum and the 40 years period. We note that the 2015 annual report of the Swedish air traffic authority does mention that ATC disturbances linked to solar radio bursts have already occurred in Sweden in 1999 and 2003, but with milder effects on the operations \citep{lfv2015}. 
 
\begin{table*}[t]
\caption{Peak flux densities in SFU for the strongest radio bursts since 2000 (peak flux greater than 50\,000 SFU at 1415~MHz) tabulated by NOAA\tablefootmark{i} (RSTN network at 1415~MHz) and by the Nobeyama Observatory\tablefootmark{ii} (1000 and 2000~MHz)}
\label{table_big_srb}
\centering
\begin{tabular}{cccc}
\hline
\hline
Date & Flux at 1000~MHz & Flux at 1415~MHz & Flux at 2000~MHz\\
\hline
2001 April 15 & N/A & 54\,000 & N/A \\
2002 April 21 & 150\,000 & 110\,000 & 9000 \\
2006 December 06 &  N/A & 139\,000\tablefootmark{a}\tablefootmark{b}& N/A \\ 
2006 December 13 & 440\,000 & 130\,000\tablefootmark{a} & 302\,000 \\
2006 December 14 &  N/A & 55\,600 & N/A \\
2011 February 15 & 46\,000 & 54\,000 & 1500 \\
2011 September 24 & N/A & 110\,000 & N/A\\
2012 March 05 & 502\,000 & 20\,000\tablefootmark{c} & 19\,000 \\
\hline
\end{tabular}
\\
\tablefoottext{i}{\url{ftp://ftp.swpc.noaa.gov/pub/warehouse/}}
\tablefoottext{ii}{\url{http://solar.nro.nao.ac.jp/norp/index.html}}
\tablefoottext{a}{Saturation limit}
\tablefoottext{b}{\citet{cliver2011} report for that event a peak flux density of $\sim 10^6$~SFU from OVSA observations between 1 and 1.6~GHz}
\tablefoottext{c}{End of observations at peak flux; probably  underestimated}
\end{table*}%

A survey of strong solar radio bursts observed since the beginning of the millennium (see Table \ref{table_big_srb}), made from lists of events from NOAA and Nobeyama Observatory, reveals at least 8 events with peak flux density greater than 50\,000~SFU, and two events peaking between 400\,000 and 500\,000~SFU at 1000~MHz. It is worth noticing that one of these two events occurred close to the minimum of cycle 23--24. 

The electron plasma frequency in flaring active regions has a range of values, and bursts like the one on November 4th 2015 can occur at neighboring frequencies, especially in a range comprising GNSS frequencies. GNSS disturbances were indeed observed on November 4th\footnote{see \url{http://gnss.be/Atmospheric_Maps/srb_event.php?date=2015-11-04}}, but to our knowledge, without consequences on related services. 
If air traffic radars are affected when the Sun is low above the horizon, ground based GNSS systems  will be more impacted when the Sun is close to the local zenith, which corresponds to the maximum of the gain of a typical GNSS antenna \citep{carrano2009}. Services like Air Traffic management, which rely both on radars and GNSS systems, like the WAAS \citep{cerruti2008} in the USA, or EGNOS in Europe could therefore be affected at any time during the day.

Table \ref{table_big_srb} shows also, like for the present event, that peak flux density can differ by several orders of magnitude in frequency bands that are relatively close (1000 and 1415~MHz). As a matter of facts, the event investigated in this study was rather unremarkable in the radio bands typically used for space weather monitoring (RSTN) and its magnitude and potential impact went unnoticed. A monitoring at 1415~MHz solely is likely, in the future, to miss potentially problematic radio bursts for air traffic radars and this alone would justify to establish world-wide, real time solar monitoring facilities covering the whole UHF band. At least, bands allocated to important services used in air traffic activities should be closely monitored, but comprehensive studies about the impact of solar radio bursts on air traffic radio services will only be possible if more technical and operational details are made available.
 
\begin{acknowledgements}
We thank Kristoffer Leer DTU Space, Lyngby, Denmark for making all personal contacts between several parties involved in the Greenland event. KLK acknowledges helpful discussions with B. Collard, B. de Courville, L. Lap\`ene, J.-Y. Prado, R. Rosso, and B.~Roturier. Research at Paris Observatory was supported by the \emph{Agence Nationale pour la Recherche} (ANR/ASTRID, DGA) project \emph{Outils radioastronomiques pour la m\'et\'eorologie de l'espace} (ORME, contract No. ANR-14-ASTR-0027). The construction of the CALLISTO spectrometer was financed by the Swiss National Science Foundation SNSF (grant nr. 20-67995.02). The Belgian solar radio instrumentation is funded by the Solar Terrestrial Center of Excellence.

The editor thanks Bryn Jones, Dale E. Gary and an anonymous referee for their assistance in evaluating this paper.
\end{acknowledgements}


\begin{thebibliography}{36}
\providecommand{\natexlab}[1]{#1}
\providecommand{\url}[1]{\texttt{#1}}
\providecommand{\urlprefix}{URL }
\providecommand{\eprint}[2][]{\url{#2}}

\bibitem[{{Aurass} et~al.(2003){Aurass}, {Klein}, {Zlotnik}, and
  {Zaitsev}}]{aurass2003}
{Aurass}, H., K.-L. {Klein}, E.~Y. {Zlotnik}, and V.~V. {Zaitsev}.
\newblock {Solar type IV burst spectral fine structures . I. Observations}.
\newblock \emph{\aap}, \textbf{410}, 1001--1010, 2003.

\bibitem[{{Bala} et~al.(2002){Bala}, {Lanzerotti}, {Gary}, and
  {Thomson}}]{bala2002}
{Bala}, B., L.~J. {Lanzerotti}, D.~E. {Gary}, and D.~J. {Thomson}.
\newblock {Noise in wireless systems produced by solar radio bursts}.
\newblock \emph{Radio Science}, \textbf{37}(2), 1018, 2002.

\bibitem[{{Benz} et~al.(2005){Benz}, {Monstein}, and {Meyer}}]{benz2005b}
{Benz}, A.~O., C.~{Monstein}, and H.~{Meyer}.
\newblock {Callisto - a new concept for solar radio spectrometers}.
\newblock \emph{\solphys}, \textbf{226}, 143--151, 2005.

\bibitem[{{Boischot}(1958)}]{boischot1958}
{Boischot}, A.
\newblock {{\'E}tude du rayonnement radio{\'e}lectrique solaire sur 169 MHz
  {\`a} l'aide d'un grand interf{\'e}rom{\`e}tre {\`a} r{\'e}seau}.
\newblock \emph{Annales d'Astrophysique}, \textbf{21}, 273, 1958.

\bibitem[{{Brueckner} et~al.(1995){Brueckner}, {Howard}, {Koomen}, {Korendyke},
  {Michels} et~al.}]{brueckner1995}
{Brueckner}, G.~E., R.~A. {Howard}, M.~J. {Koomen}, C.~M. {Korendyke}, D.~J.
  {Michels}, et~al.
\newblock {The Large Angle Spectroscopic Coronagraph (LASCO)}.
\newblock \emph{\solphys}, \textbf{162}, 357--402, 1995.

\bibitem[{{Carrano} et~al.(2009){Carrano}, {Bridgwood}, and
  {Groves}}]{carrano2009}
{Carrano}, C.~S., C.~T. {Bridgwood}, and {Groves}.
\newblock Impacts of the December 2006 solar radio bursts on the performance of
  GPS.
\newblock \emph{Radio Science}, \textbf{44}, 2009.

\bibitem[{{Cerruti} et~al.(2006){Cerruti}, {Kintner}, {Gary}, {Lanzerotti}, {de
  Paula}, and {Vo}}]{cerruti2006}
{Cerruti}, A.~P., P.~M. {Kintner}, D.~E. {Gary}, L.~J. {Lanzerotti}, E.~R. {de
  Paula}, and H.~B. {Vo}.
\newblock {Observed solar radio burst effects on GPS/Wide Area Augmentation
  System carrier-to-noise ratio}.
\newblock \emph{Space Weather}, \textbf{4}, 10,006, 2006.

\bibitem[{{Cerruti} et~al.(2008){Cerruti}, {Kintner}, {Gary}, {Mannucci},
  {Meyer}, {Doherty}, and {Coster}}]{cerruti2008}
{Cerruti}, A.~P., P.~M. {Kintner}, D.~E. {Gary}, A.~J. {Mannucci}, R.~F.
  {Meyer}, P.~{Doherty}, and A.~J. {Coster}.
\newblock {Effect of intense December 2006 solar radio bursts on GPS
  receivers}.
\newblock \emph{Space Weather}, \textbf{61}, S10D07, 2008.

\bibitem[{{Chernov}(2006)}]{chernov2006}
{Chernov}, G.~P.
\newblock {Solar Radio Bursts with Drifting Stripes in Emission and
  Absorption}.
\newblock \emph{\ssr}, \textbf{127}, 195--326, 2006.

\bibitem[{{Cliver} et~al.(2011){Cliver}, {White}, and
  {Balasubramaniam}}]{cliver2011}
{Cliver}, E.~W., S.~M. {White}, and K.~S. {Balasubramaniam}.
\newblock {The Solar Decimetric Spike Burst of 2006 December 6: Possible
  Evidence for Field-aligned Potential Drops in Post-eruption Loops}.
\newblock \emph{\apj}, \textbf{743}, 145, 2011.

\bibitem[{{Demyanov} et~al.(2012){Demyanov}, {Afraimovich}, and
  {Jin}}]{demyanov2012}
{Demyanov}, V.~V., E.~L. {Afraimovich}, and S.~{Jin}.
\newblock An evaluation of potential solar radio emission power threat on GPS
  and GLONASS performance.
\newblock \emph{GPS Solutions}, \textbf{16}, 411, 2012.

\bibitem[{{FAA}(2012)}]{faa2012}
{FAA}.
\newblock Instrument Flying Handbook.
\newblock \emph{Tech. rep.}, US Department of Transportation - Federal Aviation
  Administration,
  https://www.faa.gov/regulations\_policies/handbooks\_manuals/aviation/media/faa-h-8083-15b.pdf,
  2012.

\bibitem[{{Frequentis AG}(2007)}]{frequentis2007}
{Frequentis AG}.
\newblock B-AMC Interference Analysis and Spectrum Requirements.
\newblock \emph{Tech. rep.}, Eurocontrol,
  https://www.eurocontrol.int/sites/default/files/article/content/documents/communications/22102007-b-amc-project-deliverable-d4-v11.pdf,
  2007.

\bibitem[{{Garcia}(1994)}]{garcia1994}
{Garcia}, H.~A.
\newblock {Temperature and emission measure from GOES soft X-ray measurements}.
\newblock \emph{\solphys}, \textbf{154}, 275--308, 1994.

\bibitem[{{Hey}(1946)}]{hey1946}
{Hey}, J.~S.
\newblock {Solar Radiations in the 4-6 Metre Radio Wave-Length Band}.
\newblock \emph{\nat}, \textbf{157}, 47--48, 1946.

\bibitem[{{ICAO}(2007)}]{icao2007}
{ICAO}.
\newblock Guidance Material on Comparison of Surveillance Technologies (GMST).
\newblock \emph{Tech. rep.}, International Civil Aviation Organization,
  http://www.icao.int/RO\_APAC/Documents/edocs/cns/gmst\_technology.pdf, 2007.

\bibitem[{{Isliker} and {Benz}(1994)}]{isliker1994}
{Isliker}, H., and A.~O. {Benz}.
\newblock {Catalogue of 1-3 GHz solar flare radio emission}.
\newblock \emph{\aaps}, \textbf{104}, 1994.

\bibitem[{{Ji{\v r}i{\v c}ka} et~al.(2001){Ji{\v r}i{\v c}ka}, {Karlick{\'y}},
  {M{\'e}sz{\'a}rosov{\'a}}, and {Sn{\'{\i}}{\v z}ek}}]{jiricka2001}
{Ji{\v r}i{\v c}ka}, K., M.~{Karlick{\'y}}, H.~{M{\'e}sz{\'a}rosov{\'a}}, and
  V.~{Sn{\'{\i}}{\v z}ek}.
\newblock {Global statistics of 0.8-2.0 GHz radio bursts and fine structures
  observed during 1992-2000 by the Ond{\v r}ejov radiospectrograph}.
\newblock \emph{\aap}, \textbf{375}, 243--250, 2001.

\bibitem[{{Kennewell}(2008)}]{kennewell2008}
{Kennewell}, J.~A.
\newblock RSTN Solar Radio Telescopes (Discrete Frequency) and Data, 2008.
\newblock \urlprefix\url{www.deepsouthernskies.com/LSO/RSTN.pdf}.

\bibitem[{{Klobuchar} et~al.(1999){Klobuchar}, {Kunches}, and
  {VanDierendonck}}]{klobuchar1999}
{Klobuchar}, J.~A., J.~M. {Kunches}, and A.~J. {VanDierendonck}.
\newblock Eye on the Ionosphere: Potential Solar Radio Burst Effects on GPS
  Signal to Noise.
\newblock \emph{GPS Solutions}, \textbf{3}(2), 69--71, 1999.

\bibitem[{{Knipp} et~al.(2016){Knipp}, {Ramsay}, {Beard}, {Boright}, {Cade}
  et~al.}]{knipp2016}
{Knipp}, D.~J., A.~C. {Ramsay}, E.~D. {Beard}, A.~L. {Boright}, W.~B. {Cade},
  et~al.
\newblock {The May 1967 great storm and radio disruption event: Extreme space
  weather and extraordinary responses}.
\newblock \emph{Space Weather}, \textbf{14}, 614--633, 2016.

\bibitem[{{Lecacheux}(2000)}]{lecacheux2000}
{Lecacheux}, A.
\newblock {The Nan{\c c}ay Decameter Array: A Useful Step Towards Giant, New
  Generation Radio Telescopes for Long Wavelength Radio Astronomy}.
\newblock \emph{Washington DC American Geophysical Union Geophysical Monograph
  Series}, \textbf{119}, 321, 2000.

\bibitem[{{Lemen} et~al.(2012){Lemen}, {Title}, {Akin}, {Boerner}, {Chou}
  et~al.}]{lemen2012}
{Lemen}, J.~R., A.~M. {Title}, D.~J. {Akin}, P.~F. {Boerner}, C.~{Chou}, et~al.
\newblock {The Atmospheric Imaging Assembly (AIA) on the Solar Dynamics
  Observatory (SDO)}.
\newblock \emph{\solphys}, \textbf{275}, 17--40, 2012.

\bibitem[{{Leping} and {Zhang}(2009)}]{leping2009}
{Leping}, L., and {Zhang}.
\newblock {Statistics of flares sweeping across sunspots}.
\newblock \emph{\apjl}, \textbf{706}, L17--L21, 2009.

\bibitem[{{Luftfartsverket}(2015)}]{lfv2015}
{Luftfartsverket}.
\newblock Annual Report, 2015.
\newblock
  \urlprefix\url{https://www.lfv.se/globalassets/nyheter/nyheter-2016/eng_lfv2015_lores.pdf}.

\bibitem[{{Mercier} et~al.(1989){Mercier}, {Genova}, and
  {Aubier}}]{mercier1989}
{Mercier}, C., F.~{Genova}, and M.~G. {Aubier}.
\newblock {Radio observations of atmospheric gravity waves}.
\newblock \emph{Annales Geophysicae}, \textbf{7}, 195--202, 1989.

\bibitem[{{Micallef}(2009)}]{micallef2009}
{Micallef}, J.
\newblock Compatibility criteria and interference scenarios for SSR systems.
\newblock \emph{Tech. rep.}, Eurocontrol,
  https://www.eurocontrol.int/sites/default/files/article//content/documents/communications/24082009-lcis-c3-criteria-and-tests-v10.pdf,
  2009.

\bibitem[{{Nindos} et~al.(2008){Nindos}, {Aurass}, {Klein}, and
  {Trottet}}]{nindos2008}
{Nindos}, A., H.~{Aurass}, K.-L. {Klein}, and G.~{Trottet}.
\newblock {Radio Emission of Flares and Coronal Mass Ejections. Invited
  Review}.
\newblock \emph{\solphys}, \textbf{253}, 3--41, 2008.

\bibitem[{{Nita} et~al.(2002){Nita}, {Gary}, {Lanzerotti}, and
  {Thomson}}]{nita2002}
{Nita}, G.~M., D.~E. {Gary}, L.~J. {Lanzerotti}, and D.~J. {Thomson}.
\newblock {The Peak Flux Distribution of Solar Radio Bursts}.
\newblock \emph{\apj}, \textbf{570}, 423--438, 2002.

\bibitem[{Pedregosa et~al.(2011)Pedregosa, Varoquaux, Gramfort, Michel, Thirion
  et~al.}]{pedregosa2011}
Pedregosa, F., G.~Varoquaux, A.~Gramfort, V.~Michel, B.~Thirion, et~al.
\newblock Scikit-learn: Machine Learning in {P}ython.
\newblock \emph{Journal of Machine Learning Research}, \textbf{12}, 2825--2830,
  2011.

\bibitem[{{Pick} and {Vilmer}(2008)}]{pick2008}
{Pick}, M., and N.~{Vilmer}.
\newblock {Sixty-five years of solar radioastronomy: flares, coronal mass
  ejections and Sun Earth connection}.
\newblock \emph{\aapr}, \textbf{16}, 1--153, 2008.

\bibitem[{{R\'egnier}(2015)}]{regnier2015}
{R\'egnier}, S.
\newblock A new approach to the maser emission in the solar corona.
\newblock \emph{\aap}, \textbf{581}, A9, 2015.

\bibitem[{{Tapping}(2013)}]{tapping2013}
{Tapping}, K.~F.
\newblock {The 10.7 cm solar radio flux (F10.7)}.
\newblock \emph{Space Weather}, \textbf{11}, 394--406, 2013.

\bibitem[{{The Local}(2015)}]{tloc}
{The Local}.
\newblock Solar storm grounds Swedish air traffic, 2015.
\newblock
  \urlprefix\url{https://www.thelocal.se/20151104/solar-storm-grounds-swedish-air-traffic}.

\bibitem[{{Torii} et~al.(1979){Torii}, {Tsukiji}, {Kobayashi}, {Yoshimi},
  {Tanaka}, and {Enome}}]{torri1979}
{Torii}, C., Y.~{Tsukiji}, S.~{Kobayashi}, N.~{Yoshimi}, H.~{Tanaka}, and
  S.~{Enome}.
\newblock {Full-automatic radiopolarimeters for solar patrol at microwave
  frequencies}.
\newblock \emph{Proceedings of the Research Institute of Atmospherics, Nagoya
  University}, \textbf{26}, 129--132, 1979.

\bibitem[{{Zlotnik}(2013)}]{zlotnik2013}
{Zlotnik}, E.~Y.
\newblock {Instability of Electrons Trapped by the Coronal Magnetic Field and
  Its Evidence in the Fine Structure (Zebra Pattern) of Solar Radio Spectra}.
\newblock \emph{\solphys}, \textbf{284}, 579--588, 2013.

\end{thebibliography}
\end{document}